\documentstyle[11pt]{article}
\title{Correlation effects in ionic crystals: \\ I. The cohesive energy of MgO}
\author{Klaus Doll, Michael Dolg, Peter Fulde \\ \\
Max-Planck-Institut f\"ur Physik komplexer Systeme \\
D-01187 Dresden, Germany \\ \\  Hermann Stoll
\\ \\ Institut f\"ur Theoretische Chemie \\
        Universit\"at  Stuttgart\\
        D-70550 Stuttgart,  Germany \\}
\begin{document}
\begin{titlepage}
\maketitle
\begin{abstract}
\noindent High-level quantum-chemical calculations, using the coupled-cluster
approach and
extended one-particle basis sets, have been performed for
(Mg$^{2+}$)$_n$(O$^{2-}$)$_m$ clusters
embedded in a Madelung potential.
The results of these calculations are used for setting up an incremental
expansion for
the correlation energy of bulk MgO. This way, $\sim$ 96\% of the experimental
cohesive
energy of the MgO crystal is recovered.
It is shown that only $\sim$ 60\% of the correlation contribution to the
cohesive
energy is of intra-ionic origin, the remaining part being caused by van der
Waals-like
inter-ionic excitations.
\end{abstract}
accepted by Phys. Rev. B
\end{titlepage}

\section{Introduction}

While the density-functional (DFT) method, with its long tradition in
solid-state physics,
is getting wide acceptance in the field of quantum chemistry nowadays, there
are also attempts to the
reverse, i.e., making use of the traditional quantum-chemical Hartree-Fock (HF)
and
configuration-interaction (CI) methods not only for molecular but also for
solid-state
applications. The Torino group of Pisani and co-workers, e.g., devised an
ab-initio HF scheme
for solids [1], which has successfully been applied to a broad range of
(mostly) covalently
bonded and ionic solids, within the past five years. A main asset of the HF
scheme is the
availability of a well-defined wavefunction, which may not only be used for
extracting properties
but also as a starting-point for systematically including electron-correlation
effects.
Such effects, which are only implicitly accounted for  in density-functional
methods, often
have a strong influence on physical observables, in molecules as well as in
solids.
Several suggestions have been made how to explicitly include electron
correlation in solids, among
them the Quantum Monte-Carlo (QMC) approach [2], the Local Ansatz (LA) [3], and
the method
of local increments [4] (which may be considered as a variant of the LA); in
QMC
the HF wavefunction is globally corrected for electron-correlation effects by
multiplying it with a factor containing
inter-electronic coordinates (Jastrow factor); the latter two methods rely on
applying selected
local excitation operators to the HF wavefunction and thus have a rather close
connection to
traditional quantum-chemical post-HF methods.

The number of test examples is still rather
limited with all three solid-state correlation schemes, and is mainly
restricted to
semiconductors so far. For ionic insulators where quantum-chemical methods
would seem to be
most suitable and easily advocated, much work indeed has been devoted to
correlation effects on band structures (cf.\ e.g.\
[5a,b]), but only few studies refer to cohesive energies (cf.\ e.g.\ [5c]), and
only a single application of the post-HF schemes
mentioned in the last paragraph exists to our knowledge (NiO with QMC [5d]).
This does not mean that such applications to
ground-state correlation effects are without challenge. For MgO, the system to
be dealt
with in this paper, HF calculations [6] yield a lattice constant which is in
agreement
with experiment to $\sim$ 0.01 \AA, but the correlation contribution to the
cohesive
energy is significant ($\sim$ 3 eV, nearly half as large as the HF value). The
local-density
approximation (LDA) of DFT does not a good job here either: an overbinding
results which is
more than two times as large as the correlation contribution to the lattice
energy [7], and invocation of gradient-corrected
functionals is indispensable for obtaining reasonable results [6b], cf.\ Sect.\
3.5 below.
The situation would not seem too complicated, nevertheless, if the effect could
be explained
just by adding correlation contributions of individual ions; in fact, such a
suggestion has repeatedly
been made in literature, cf.\ Refs.\ [6a,17c]. However, the O$^{2-}$ ion, one
of the building blocks of the MgO
crystal, is not stable as a free entity, and an accurate determination of the
correlation
energy of this highly fluctuating and easily polarizable ion in its crystal
surroundings
is not expected to be an easy task; the more so, since already for the
determination of the
electron affinity of the free O atom high-level quantum-chemical correlation
methods
are required [8]. Moreover, as we shall show below, intra-ionic interactions
contribute
only with $\sim$ 60\% to the total correlation effect on the bulk cohesive
energy, and
van der Waals-like inter-ionic excitations play an important role here.

This paper is the first in a series devoted to the application of the method of
local
increments to ionic solids; it shows, at the example of MgO, how to set up an
incremental expansion
of the bulk correlation energy using information from quantum-chemical
calculations on finite clusters only. The organization of the paper is as
follows:
in Section 2, computational details are given for the applied quantum-chemical
methods,
and test calculations are performed for the first two ionization potentials of
the Mg atom,
for the electron affinity of the O atom, and for spectroscopic properties of
the MgO molecule.
The method of local increments is briefly described in
Section 3; correlation-energy increments are evaluated from calculations on
(Mg$^{2+}$)$_n$(O$^{2-}$)$_m$
clusters, and the incremental expansion for the total correlation energy of
bulk MgO is
discussed.
Conclusions follow in Section 4.

\section{Test calculations}

Our first test concerns the electron affinity, EA, of the oxygen atom. We
performed separate
calculations for the ground states of O and O$^-$, respectively, using various
single- and multi-reference quantum-chemical configuration interaction methods
for treating many-body
correlation effects [9]. The one-particle basis sets employed have been
taken from the series of correlation-consistent (augmented) polarized Gaussian
basis sets
of Dunning and co-workers [10,8a]. The ab-initio program package MOLPRO [11]
has been used in these
and all of the following calculations of the present paper. The results for
EA(O) are collected
in Table 1.
It is seen that the best single- and multi-reference methods (coupled-cluster
with
single and double excitations and perturbative inclusion of triples (CCSD(T))
and
2$s$ - 2$p$ active space; multi-reference averaged coupled pair functional with
single and double
excitations (MR-ACPF) and 2$s$ - 3$p$ active space) yield quite similar results
(1.40 eV), which
differ from the experimental value by only 0.06 eV. Effects of the one-particle
basis set are significant,
even at the stage of including $g$ functions, and probably are responsible for
the major part of the
remaining deviation from experiment. In the (Mg$^{2+}$)$_n$(O$^{2-}$)$_m$
cluster calculations
to be described in Sect.\ 3, we could only afford
the valence triple-zeta [$5s4p3d2f$] basis set, at the single-reference level;
the concomitant differential errors can be
estimated to about 0.1 eV per O atom and (added) electron.

The next test deals with the ionization potentials, IP, of the magnesium atom
(Mg $\rightarrow$ Mg$^+$,
Mg$^+$ $\rightarrow$ Mg$^{2+}$). A difficulty is encountered here, since not
only the correlation energy
of the valence (3$s^2$) electron pair has to be accounted for but also
core-valence correlation effects
are non-negligible: the latter contribute with 0.3 eV to the Mg$^+$
$\rightarrow$ Mg$^{2+}$ IP, e.g..
For accurately describing these effects explicitly, a high computational effort
is needed in quantum-chemical
ab-initio calculations. In (all-electron) calculations with a basis set of
medium quality
(($12s9p1d$)/[$5s4p1d$] [12]) errors of 0.12 and 0.21 eV remain for the two
IPs, and even a very large
uncontracted ($20s15p6d3f$) basis [13]
still yields deviations from experiment of 0.03 and 0.04 eV in CCSD(T)
calculations.
Without loss of accuracy, however, the computational effort can be effectively
reduced [14,15] by simulating the
Mg$^{2+}$ core by a pseudopotential (PP) which describes core-valence
interaction at the HF level,
in conjunction with a core-polarization potential (CPP) which accounts for
core-valence
correlation effects. Using these methods, very good agreement with experiment
is obtained, at the correlated
level, cf.\ the results of Table 2. (Only ACPF data are given, in the Table,
since for a
two-electron system all the correlation methods of Table 1 (CI, ACPF, CCSD)
coincide.)
For the cluster calculations of Sect.\ 3, we adopt the PP+CPP description,
together with the energy-optimized
($4s4p$) valence basis set; the concomitant differential errors can be
estimated to about 0.02 eV per Mg atom
and (removed) electron.

The last test of our quantum-chemical arsenal of methods was performed for
the MgO molecule. The basis sets applied here
are the same as those used in the next section for
(Mg$^{2+}$)$_n$(O$^{2-}$)$_m$ clusters.
The results for bond length, dissociation energy and vibrational frequency are
compiled in Table 3.
It is seen that excellent agreement with experiment (to 0.02 \AA, 0.1 eV, 10
cm$^{-1}$(1\%)) is obtained
at the (single-reference) CCSD(T) level. At the multi-reference level (without
triples), the agreement
is slightly less good, but this could certainly be improved upon by enlarging
the 2-configuration reference
space which was chosen in our calculations.

\section{Local increments}
\subsection{Methodological aspects}

The main idea to be discussed here is the possibility to extract information
from calculations
on finite clusters and to transfer it to the infinite crystal. Such a transfer
would certainly {\em not}
be a good idea for {\em global} cluster properties (cohesive energy, ionization
potential, etc.) --
it only makes sense for {\em local} quantities. Now, localized orbitals are
entities which can be defined
within ionic crystals as well as within clusters of these materials. Moreover,
electron correlation in or
between such orbitals is a local effect. Therefore, if we prepare localized
orbitals in the interior of a
cluster (in a sufficiently solid-like environment) and if we calculate
correlation energies involving
these orbitals, we can hope to obtain transferable quantities.

Of course, there is no reason to expect that these quantities would be additive
in the solid.
If we separately calculate, e.g., the pieces of correlation energy due to
orbitals
localized at ionic positions $A$, $B$, $C$, ...
\begin{equation}
\epsilon(A), \epsilon(B), \epsilon(C), ...
\end{equation}
the correlation energy of the common orbital system of $AB$ (or $AC$, $BC$,
...) will deviate in general from the
sum of constituents, due to inter-ionic interactions, and we can define
non-additivity corrections such as
\begin{equation}
\Delta\epsilon(AB) =  \epsilon(AB) - \epsilon(A) - \epsilon(B).
\end{equation}
Again, the next larger systems of three ions, $ABC$, ..., will have correlation
energies slightly different from
the sum of the constituents plus the two-body non-additivity corrections, and
this gives rise to three-body
increments
\begin{eqnarray}
\Delta\epsilon(ABC) = & \epsilon(ABC) -   [\epsilon(A) + \epsilon(B) +
\epsilon(C)] - \nonumber  \\
                                                 &  [\Delta\epsilon(AB) +
\Delta\epsilon(AC) + \Delta\epsilon(BC)].
\end{eqnarray}
Similar definitions apply, in principle, to higher-body increments.

If we now make use of all these quantities, i.e., the intra-ionic correlation
energies and the various
inter-ionic correction terms, and multiply them with weight factors appropriate
for the solid,
we can hope to get a meaningful incremental expansion of the correlation energy
per unit cell of the
infinite crystal:
\begin{equation}
\epsilon_{\rm bulk} = \sum_A \epsilon(A) + \frac{1}{2} \sum_{A,B}
\Delta\epsilon(AB) + \frac{1}{3!} \sum_{A,B,C}
\Delta\epsilon(ABC) + ...
\end{equation}
In Ref.\ [4d] we have shown that this equation can be formally derived, under
appropriate approximations,
from an expression for the correlation energy of an infinite system.

Let us discuss now the assumptions implicit in this approach more closely for
the case of the MgO crystal.
MgO is generally considered as a nearly perfect ionic crystal consisting of
Mg$^{2+}$ and O$^{2-}$ ions [17];
the question of a quantitative measure for the ionicity of MgO has been
addressed only recently by Bagus and
co-workers [17c,d], and in careful
studies using various criteria the ionic charges have been shown to deviate
from $\pm$2 by $<$0.1 only.
Thus, the attribution of localized orbitals to ionic positions made above seems
to be a valid assumption.
But even if there were some degree of covalency and/or some tendency for
delocalization in MgO, this would not
invalidate our approach. In fact, the first applications of the method of local
increments were made for
covalently bonded crystals (diamond, silicon [4a,b]), and even for the
$\pi$-system of graphite which
according to usual classifications is considered as completely delocalized the
method has been shown to
yield meaningful results [4c].

Secondly, the determination of increments for non-additive inter-ionic
correlation contributions in Eqs.\ 2
and 3 makes sense only, if the number of non-negligible increments is small,
i.e., if the
$\Delta\epsilon(AB)$ rapidly decrease with increasing distance of the ions and
if the three-body terms
$\Delta\epsilon(ABC)$ are significantly smaller than the two-body ones making
the use of four-body contributions
obsolete. A necessary pre-requisite for satisfying these conditions is the use
of a size-extensive correlation
method for calculating the increments. This excludes, for instance, the
(variational) configuration-interaction
method with single and double excitations (CISD), since it does not scale
linearly with $n$ for a system of
non-interacting atoms $A_n$. On the other hand, correlation methods of the
coupled-cluster variety such as
those discussed in the last section (ACPF, CCSD, CCSD(T)) do  have this
property. (These are benchmark methods of increasing
complexity widely used in quantum chemistry; we display results derived from
all of them in the following Tables, in order to
monitor convergence with respect to the many-particle basis set used.) When
applying such a method,
$\Delta\epsilon(AB)$ should indeed rapidly decrease with increasing $AB$
distance,
since for non-overlapping pairs of ions only
van der Waals-like correlation effects ($\sim 1/r^6$) become effective.
Furthermore, three-body terms
can be expected to be significantly smaller than two-body terms, since
two-electron excitations, involving a
pair of orbitals at most, are known to dominate correlation effects.

A final assumption underlying our approach is that of the transferability of
localized orbitals from clusters to the
bulk which was mentioned right at the beginning of this subsection.
Its fulfillment depends, of course, on the preparation of the
clusters. A free O$^{2-}$ ion, e.g., would be unstable, and it is essential,
therefore, to put this ion in a cage of
Mg$^{2+}$ ions in order to stabilize it, and to simulate the Madelung potential
of the surrounding ions in order to
provide the correct field near the O nucleus. More details on cluster
preparation and
transferability tests will be given in the following subsections, where the
determination of individual
increments is discussed.

\subsection{Intra-ionic correlation}

The first increment to be calculated is the correlation energy which can be
locally attributed to an O$^{2-}$
ion in crystal surroundings of Mg$^{2+}$ and other O$^{2-}$ ions. As already
mentioned, a realistic modelling of
the crystal surroundings is essential, since otherwise the O$^{2-}$ ion would
not be stable at all. Fortunately,
stabilization can be achieved in a both very simple and efficient way, by
simulating the Pauli repulsion of the
6 nearest-neighbour Mg$^{2+}$ ions by means of pseudopotentials; we used the
same energy-consistent
pseudopotentials here as were used for the treatment of the Mg atom in the
calculations of Sect.\ 2.
For representing the crystal environment of the resulting 7-atom cluster, a
Madelung approximation was made:
336 ions surrounding this cluster in a cube of 7x7x7 ions were simulated by
point charges $\pm$2 (with charges
at the surface planes/edges/corners reduced by factors 2/4/8, respectively).
Here and in the following, the
experimental bulk lattice constant ($r_{\rm MgO}$ = 2.105 \AA) was adopted.
Employing the [$5s4p3d2f$]
basis set, which was already used
for O and O$^-$, for the description of the O$^{2-}$ orbitals, too,  we obtain
the differential
correlation-energy contributions, $\Delta\epsilon$(O$^{2-}$) =
$\epsilon$(O$^{2-}$)-$\epsilon$(O), to the affinity of the extra
electrons in crystal O$^{2-}$ which are listed in Table 4.

A first point to make is that at all levels of approximation
$\Delta\epsilon$(O$^{2-}$) comes out considerably
smaller than one would expect from a simple linear scaling of
$\Delta\epsilon$(O$^-$) values (2.77 vs.\ 1.86 eV,
at the CCSD(T) level, cf.\ Table 1); such a linear scaling, which approximately
works for the iso-electronic
systems, $\Delta\epsilon$(Ne) and $\Delta\epsilon$(Ne$^-$), [18] probably fails
for O$^{2-}$ due to the increased
spacing of excited-state levels, when compressing the O$^{2-}$ charge density
in the (Mg$^{2+}$)$_6$ cage.
When comparing individual $\Delta\epsilon$(O$^{2-}$) values in Table 4, we
observe that
in our single-reference calculations
(active space 2$s$ - 2$p$), the effect of single and double excitations is
quite similar for ACPF and
CCSD, while the inclusion of triples in CCSD(T) yields an increase  by another
5\%. Thus, the
effect of triples is of less (relative) importance than in free O$^-$ but is
still non-negligible. As in the
case of O$^-$, we checked that enlarging the active space (to 2$s$ - 3$p$) in
the ACPF calculations (MRACPF)
leads to a result numerically nearly identical to CCSD(T). Moreover, we tested
the influence of an increase of
the basis set ([$5s4p3d2f$] $\rightarrow$ [$6s5p4d3f2g$], cf.\ Table 1);
at the CCSD(T) level, the correlation-energy increment
changes by -0.008 a.u. (-0.2 eV), in line with our estimate given in Sect.\ 2.
We also tried increasing the [$5s4p3d2f$] basis set by adding off-center
functions (the ($4s4p$) sets of Sect.\ 2 at the
positions of the Mg$^{2+}$ ions); the correlation-energy change of -0.005 a.u.
is somewhat
smaller here because not all components of the higher polarization functions
($f$, $g$)
at the oxygen site cannot be simulated this way.
A last test concerns the influence of the Madelung field. Leaving
out all of the point-charges and performing the calculation with the bare
(O$^{2-}$)(Mg$^{2+}$)$_6$ cluster
leads to quite negligible correlation-energy changes of $\leq 4\cdot 10^{-5}$
a.u. only, at all theoretical
levels; this underlines the notion of electron correlation being a local
effect. Summarizing,
lack of completeness of the one-particle basis set seems to be the largest
source of error in the O$^{2-}$ results
listed in Table 4, and the order of magnitude of the resulting error for the
MgO cohesive energy per unit cell
(with respect to separated neutral atoms) can be assessed to  0.2 ...\ 0.3 eV.

Let us next consider the Mg $\rightarrow$ Mg$^{2+}$ correlation-energy
increment. This increment
can directly be evaluated
using the atomic calculations described in Sect.\ 2. This is so, since the
Madelung effect identically
vanishes here, when Mg$^{2+}$ is described by a bare pseudopotential.
We checked this point by performing all-electron calculations for Mg$^{2+}$
with and without Madelung field:
the non-frozen-core effect obtained thereby is of the order of $1\cdot 10^{-6}$
a.u. with our largest basis set.
Thus, the pseudopotential approximation is certainly valid here.

Adding the Mg$^{2+}$ and O$^{2-}$ correlation-energy increments of Table 4, we
get, at the highest theoretical
level (CCSD(T)), a correlation contribution of 0.0547 a.u. to the bulk cohesive
energy, $E_{\rm coh}$,
per (primitive) unit cell of the MgO
crystal. The experimental cohesive energy corrected for zero-point energy is
known to be 0.3841 a.u. [19]; the most
recent (and probably best) HF value is 0.2762 a.u. [6c]. This yields an
'experimental' correlation contribution
to $E_{\rm coh}$ of 0.108 a.u. which is just about the double of the
intra-ionic value calculated so far. Thus, it
is clear that inter-ionic contributions to be dealt with in the next subsection
play an important role.

\subsection{Two-Body Corrections}

In this subsection, non-additivity corrections are determined, which arise when
simultaneously correlating two ions in a
cluster.

Let us first consider here the interaction of ions with charges of opposite
sign, i.e.\ Mg$^{2+}$ and O$^{2-}$,
which are next neighbours in the crystal. Using the same cluster as in the
previous subsection when determining
the intra-ionic correlation energy of O$^{2-}$ (O$^{2-}$ plus 6 surrounding
Mg$^{2+}$ plus 336 point charges)
and adding a core-polarization potential at one of the Mg$^{2+}$ neighbours of
the central O$^{2-}$ ion, we
obtain an inter-ionic core-valence correlation contribution (cf.\ Table 5)
which is due to the dynamic
polarization of the Mg$^{2+}$ core by the O$^{2-}$ valence electrons and which
was clearly absent in the free
Mg$^{2+}$ ion. Although the resulting value for the increment turns out to be
considerably smaller than the
intra-ionic correlation contributions of Table 4, its effect on the cohesive
energy of MgO is by no means
negligible, due to the large weight factor.

In order to study the convergence of the correlation-energy
increments with increasing distance of the ions, we replaced, in the cluster
described above, one of the
Madelung charges (at positions of Mg$^{2+}$ ions successively more distant from
the central O$^{2-}$ ion) by a
pseudopotential and evaluated the influence of core-polarization. A rapid
decrease with $r_{\rm MgO}$ is
observed, with the fourth-nearest neighbour MgO increment already approaching
the numerical noise in our
calculations.

Turning now to the increments related to pairs of ions of the same kind, we can
safely neglect Mg$^{2+}$-Mg$^{2+}$
interactions. The correlation-energy contributions are exactly zero, at the
pseudopotential plus
core-polarization level. All-electron test calculations yield very small values
around $\sim 2\cdot 10^{-5}$ a.u.
for a nearest-neighbour pair of Mg$^{2+}$ ions in the crystal.

Far more important are interactions between O$^{2-}$
ions with their diffuse, fluctuating charge distributions. For the increment
between nearest neighbouring
O$^{2-}$ ions, we used a cluster with 448 ions, where two central O$^{2-}$ ions
were treated explicitly, while
the 10 next Mg$^{2+}$ neighbours of these ions were simulated by
pseudopotentials
and the remaining ions were represented
by point charges. The non-additivity correction with respect to two single
O$^{2-}$ ions (cf.\ Table 5) turns
out to be of the same order of magnitude as the MgO increment. The greater
strength of interaction in the O-O
case compared to the Mg-O one (larger polarizability of O$^{2-}$ compared to
Mg$^{2+}$) is effectively
compensated by the enhanced ion distance.

Results for O$^{2-}$-O$^{2-}$ increments between 2$^{nd}$ through
5$^{th}$ neighbours are also given in Table 5. The decrease with increasing ion
distance is not quite as rapid
as that for the Mg-O increments. By multiplying the O-O increments with
$r^{-6}$, one can easily check that a
van der Waals-like behaviour is approached for large $r$, and one can use the
resulting van der Waals constant
to obtain an estimate of the neglected increments beyond 5$^{th}$ neighbours;
this estimate which is $\sim 3\cdot 10^{-4}$ a.u. (including
appropriate weight factors) should be considered as an error bar for the
truncation of the incremental expansion
of the MgO cohesive energy in our calculations.

Adding up all the two-body increments which have been determined in this work,
we find (cf.\ last row in Table
5) that the inter-ionic two-body correlation contribution to the cohesive
energy of MgO is of comparable
magnitude as the intra-ionic one. Thus, the important conclusion may be drawn
that even in a (nearly purely)
ionic crystal a mean-field description of inter-ionic interactions
(electrostatic attraction, closed-shell
repulsion) is not sufficient.

\subsection{Three-Body Corrections}

Let us complete now the information necessary for setting up the incremental
expansion of the correlation energy
of bulk MgO, by evaluating the most important non-additivity corrections
involving three ions. These corrections
are obtained for triples with at least two pairs of ions being nearest
neighbours of their respective species,
using Eq.\ 3 (cf.\ Sect.\ 3.1). The numerical results are listed in Table 6. It
is seen that the largest 3-body
corrections are smaller by nearly two orders of magnitude compared to
the leading two-body ones, thus indicating a rapid
convergency of the many-body expansion with respect to the number of atoms
included; the total three-body
contribution to the correlation piece of the bulk cohesive energy, $E_{\rm
coh}$ is $\sim$ 2\% of the two-body
part and of opposite sign.

\subsection{Incremental Expansion}

In Table 7, the sum of local correlation-energy increments to the
cohesive-energy, $E_{\rm coh}$, of MgO
(with respect to separated neutral atoms) is
compared to the difference of experimental and HF values for $E_{\rm coh}$. Our
calculated values amount to
between $\sim$ 80 and $\sim$ 85 \%, depending on the correlation method
applied, of the experimental value. The inclusion of
triple excitations in the correlation method seems to be significant for
describing the large fluctuating O$^{2-}$
ions. A major part of the remaining discrepancy to experiment is probably due
to deficiencies of the one-particle
basis set: as discussed in Sect.\ 3.2, extension of the O$^{2-}$ basis set to
include $g$ functions, for the
evaluation of the intra-ionic contribution, already reduces the error by a
factor of 2.

A comparison to related theoretical results is possible at the
density-functional level.
A correlation-energy functional including gradient corrections (Perdew 91)
yields a $\Delta$E$_{\rm coh}
$ value of 0.087 a.u.\ [6b], only slightly inferior to our CCSD(T) one. Further
density-functional
results for lattice energies have been reported in Ref.\ [7]; these results
have been determined for fixed HF
densities, and the reference of the lattice energies is to Mg$^{2+}$ + O$^-$ +
$e^-$.
As already mentioned in Sect.\ 1, the LDA
value is much too high (by 0.069 a.u.); the results with the Perdew-91
correlation-energy functional on top of the
HF exchange is much better (error 0.014 a.u.), and the same may be said for a
xc-DFT treatment with the Becke
and Perdew-91 gradient corrections for exchange and correlation, respectively
(error 0.011 a.u.).
Using the Mg$^{2+}$ and O$^-$ data of Tables 1 and 2, our present calculations
lead, at the CCSD(T) level, to
a value of 0.071 a.u. for the correlation contribution to the lattice energy;
adding this value to the HF
result of Ref.\ [7], a deviation from experiment of 0.009 a.u. is obtained,
which is similar to that of
the gradient-corrected DFT ones. Note, however, that this comparison must be
taken with a grain of salt:
the HF calculation of Ref.\ [7] is certainly not too accurate, since a
relatively small basis set was used.
Otherwise it would be difficult to explain why our correlation contribution to
the lattice energy should be
too {\em large}, although all possible sources of errors (discussed in Sects.\
3.2 and 3.3) point to the
opposite direction.

\section{Conclusion}
The correlation energy of the MgO crystal can be cast into a rapidly convergent
expansion in terms of local
increments which may be derived from finite-cluster calculations. One-particle
basis sets of triple-zeta quality
with up to $f$ functions at the positions of the O atoms and high-level
quantum-chemical correlation methods
(CCSD(T)) are necessary for obtaining $\sim$ 85\% of the correlation
contribution to the bulk cohesive energy.
About one half of this contribution can be attributed to intra-ionic
interactions, the rest is due -- to about
equal parts -- to dynamic polarization of the Mg$^{2+}$ cores by the O$^{2-}$
ions
and to van der Waals-like O$^{2-}$-O$^{2-}$ interactions. \\
Although numerical results from quantum-chemical calculations of this type are
not necessarily superior to DFT ones, they provide
additional physical insight into the sources of correlation contributions to
solid-state properties.

\section*{Acknowledgments}
We are grateful to Prof.\ H.-J.\ Werner (Stuttgart) for providing the program
package MOLPRO.

\newpage
\section*{References}
\begin{enumerate}
\item R. Dovesi, C. Pisani, and C. Roetti, Int. J. Quantum Chem. {\bf 17}, 517
(1980);
C. Pisani, R. Dovesi, and C. Roetti, Lecture Notes in Chemistry, vol. 48
(Springer, Berlin, 1988)
\item S. Fahy, X.W. Wang, and S.G. Louie, Phys. Rev. B {\bf 42}, 3503 (1990);
X.P. Li, D.M. Ceperley, and R.M. Martin, Phys. Rev. B {\bf 44}, 10929 (1991)
\item G. Stollhoff and P. Fulde, J. Chem. Phys. {\bf 73}, 4548 (1980);
P. Fulde in {\em Electron Correlations in Molecules and Solids}, Springer
Series in Solid State
Sciences, vol. 100 (Springer, Berlin, 1993)
\item H. Stoll, Chem. Phys. Lett. {\bf 191}, 548 (1992); H. Stoll, Phys. Rev. B
{\bf 46}, 6700 (1992);
H. Stoll, J. Chem. Phys. {\bf 97}, 8449 (1992); B. Paulus, P. Fulde, and H.
Stoll, Phys. Rev. B (accepted)
\item R. Pandey, J.E. Jaffe, and A.B. Kunz, Phys. Rev. B {\bf 43}, 9228 (1991);
B.H. Brandow, Adv. Phys. {\bf 26}, 651 (1977); J.M. Recio, R. Pandey, A.
Ayuela, and A.B. Kunz, J. Chem. Phys. {\bf 98}, 4783
(1993); S. Tanaka, J. Phys. Soc. Jpn. {\bf 62}, 2112 (1993)
\item M. Caus\`a, R. Dovesi, C. Pisani, and C. Roetti, Phys. Rev. B {\bf 33},
1308 (1986);
R. Dovesi, C. Roetti, C. Freyria-Fava, E. Apr\`a, V.R. Saunders, and N.M.
Harrison, Philos. Trans. R. Soc.
London Ser. A {\bf 341}, 203 (1992); M. Catti, G. Valerio, R. Dovesi, and M.
Caus\`a, Phys. Rev. B {\bf 49}, 14179
(1994)
\item M. Caus\`a and A. Zupan, Chem. Phys. Lett. {\bf 220}, 145 (1994)
\item R.A. Kendall, T.H. Dunning, Jr., and R.J. Harrison, J. Chem. Phys. {\bf
96}, 6796 (1992);
D.L. Strout and G.E. Scuseria, J. Chem. Phys. {\bf 96}, 9025 (1992)
\item For a survey of quantum-chemical ab-initio methods, cf.\ e.g.\
R.J. Bartlett and J.F. Stanton in {\em Reviews in Computational Chemistry},
vol. 5, edited by K.B.
Lipkowitz and D.B. Boyd (VCH, New York, 1994)
\item T.H. Dunning, Jr., J. Chem. Phys. {\bf 90}, 1007 (1989)
\item MOLPRO is a package of {\em ab initio} programs written by H.-J. Werner
and P.J. Knowles, with
contributions from J. Alml\"of, R.D. Amos, M.J.O. Deegan, S.T. Elbert, C.
Hampel, W. Meyer, K. Peterson,
R. Pitzer, A.J. Stone, and P.R. Taylor;
H.-J. Werner and P.J. Knowles, J. Chem. Phys. {\bf 82}, 5053 (1985);
P.J. Knowles and H.-J. Werner, Chem. Phys. Lett. {\bf 115}, 259 (1985);
H.-J. Werner and P.J. Knowles, J. Chem. Phys. {\bf 89}, 5803 (1988);
P.J. Knowles and H.-J. Werner, Chem. Phys. Lett. {\bf 145}, 514 (1988);
H.-J. Werner and P.J. Knowles, Theor. Chim. Acta {\bf 78}, 175 (1990);
C. Hampel, K. Peterson, and H.-J. Werner, Chem. Phys. Lett. {\bf 190}, 1
(1992);
P.J. Knowles, C. Hampel, and H.-J. Werner, J. Chem. Phys. {\bf 99}, 5219 (1993)
\item A.D. McLean and G.S. Chandler, J. Chem. Phys. {\bf 72}, 5639 (1980)
\item ($20s12p$) basis set from H. Partridge, J. Chem. Phys. {\bf 87}, 6643
(1987),
supplemented by 3 diffuse $p$ functions and a $6d3f$ polarization set;
$\sim$50\% of the deviation from experiment,
in our CCSD(T) calculations for the first two Mg IP with this basis, are of
relativistic origin.
\item P. Fuentealba, L. v. Szentp\'aly, H. Preuss, and H. Stoll, J. Phys. B
{\bf 18}, 1287 (1985)
\item W. M\"uller, J. Flesch, and W. Meyer, J. Chem. Phys. {\bf 80}, 3297
(1984);
P. Fuentealba, H. Preuss, H. Stoll, and L. v. Szentp\'aly, Chem. Phys. Lett.
{\bf 89}, 418 (1982);
L. v. Szentp\'aly, P. Fuentealba, H. Preuss, and H. Stoll, Chem. Phys. Lett.
{\bf 93}, 555 (1983);
H. Stoll, P. Fuentealba, M. Dolg, J. Flad, L. v. Szentp\'aly, and H. Preuss, J.
Chem. Phys. {\bf 79}, 5532
(1983)
\item K.P. Huber and G. Herzberg, {\em Molecular Spectra and Molecular
Structure: IV. Constants of Diatomic
Molecules} (Van Nostrand, New York, 1979);
L. Operti, D.C. Tews, T.J. MacMahon, and B.S. Freiser, J. Am. Chem. Soc. {\bf
111}, 9152 (1989)
\item S. Sasaki, K. Fujino, and Y. Takeuchi, Proc. Japan. Acad. B {\bf 55}, 43
(1979);
Y.N. Xu and  W.Y. Ching, Phys. Rev. B {\bf 43}, 4461 (1991);
F. Illas, A. Lorda, J. Rubio, J.B. Torrance, and P.S. Bagus, J. Chem. Phys.
{\bf 99}, 389 (1993);
G. Pacchioni, C. Sousa, F. Illas, F. Parmigiani, and P.S. Bagus, Phys. Rev. B
{\bf 48}, 11573 (1993);
C. Sousa, F. Illas, C. Bo, and J.M. Poblet, Chem. Phys. Lett. {\bf 215}, 97
(1993);
R. Orlando, R. Dovesi, C. Roetti, and V.R. Saunders, Chem. Phys. Lett. {\bf
228}, 225 (1994)
\item E.R. Davidson, S.A. Hagstrom, S.J. Chakravorty, V.M. Umar, and Ch. Froese
Fischer, Phys. Rev. A {\bf 44}, 7071
(1991)
\item R.C. Weast, ed., {\em CRC Handbook of Chemistry and Physics} (CRC Press,
Boca Raton, 1987)
\end{enumerate}

\newpage
\begin{table}
\caption{\label{1}Electron affinity (eV) for the O atom at various theoretical
levels (RHF: restricted Hartree-Fock
self-consistent-field, CASSCF: complete active space self-consistent-field,
(MR)CI: (multi-reference)
configuration interaction, (MR)ACPF: (multi-reference) averaged coupled pair
functional, CCSD(T)
coupled-cluster with single and double substitutions (and perturbative
correction for triples);
the reference configurations comprise
all possible distributions of electrons within the space of active orbitals.}
\vspace{1cm}
\begin{minipage}{5in}
\begin{center}
\begin{tabular}{|c|ccc|}
\hline
basis set & [$5s4p3d2f$]\footnote{Correlation-consistent augmented polarized
valence triple- and quadruple-zeta
basis sets from Refs.\ [10,8a], contracted from ($11s6p3d2f$) and
($13s7p4d3f2g$) Gaussian primitives, respectively.
} & [$6s5p4d3f2g$]$^a$ & [$6s5p4d3f2g$]$^a$ \\
active space & $2s-2p$ & $2s-2p$ & $2s-3p$ \\
\hline
RHF & -0.53 & -0.54 & -0.54 \\
CASSCF & -0.53 & -0.53 & 0.77 \\
(MR)CI & 1.02 & 1.07 & 1.36 \\
(MR)ACPF & 1.23 & 1.28 & 1.40 \\
CCSD & 1.18 & 1.24 & \\
CCSD(T) & 1.33 & 1.40 & \\
\hline
exptl. & \multicolumn{3}{|c|}{1.46} \\
\hline
\end{tabular}
\end{center}
\end{minipage}
\end{table}

\newpage
\begin{table}
\begin{minipage}{5in}
\begin{center}
\caption{\label{2}Ionization potentials
Mg$\rightarrow$Mg$^+$/Mg$^+$$\rightarrow$Mg$^{2+}$ of the
magnesium atom (in eV), from calculations using a 2-valence-electron
pseudopotential (PP);
in the (single-reference) ACPF calculations a core-polarization potential
accounts
for core-valence correlation.}
\vspace{5mm}
\begin{tabular}{|c|cc|}
\hline
\multicolumn{1}{|c|}{basis set\footnote{Valence basis set optimized for the PP;
exponents of Gaussian primitives are
$s$: 2.4257193, 0.8226250, 0.1077490, 0.0394850; $p$: 0.7690470, 0.1886750,
0.0751010, 0.0294970;
$d$: 0.2; the basis set is used without contraction.}} & ($4s4p$) & ($4s4p1d$)
\\
\hline
RHF\footnote{For explanation of acronyms, cf.\ Table \ref{1}.} & 6.63/14.75 &
6.63/14.75 \\
ACPF$^b$ & 7.63/15.02 & 7.65/15.02 \\
\hline
exptl. & \multicolumn{2}{|c|}{7.65/15.04} \\
\hline
\end{tabular}
\end{center}
\end{minipage}
\end{table}

\newpage
\begin{table}
\begin{center}
\begin{minipage}{5in}
\caption{\label{3}Bond length $R_e$ (\AA), dissociation energy $D_e$ (eV) and
vibrational frequency
$\omega_e$ (cm$^{-1}$) of the MgO
molecule, evaluated at various theoretical levels. A Mg$^{2+}$ pseudopotential
was used, and, at the
correlated levels, a core-polarization potential was added. In the
multi-reference calculations, a
5-orbital (Mg 3s, O 2s,2p) active space was chosen.}
\vspace{5mm}
\begin{tabular}{|c|ccc|}
\hline
 &  $R_e$ & \multicolumn{1}{c}{$D_e$} & $\omega_e$ \\
\hline
RHF\footnote{For explanation of the acronyms,
cf. Table 1.
Basis sets: O [$5s4p3d2f$]; Mg ($4s4p$).}
& 1.715 & - & 794 \\
MRCI$^a$ & 1.750 & 2.01 & 751 \\
MRACPF$^a$ & 1.751 & 2.29 & 755  \\
CCSD$^a$ & 1.744 & 1.96 & 746 \\
CCSD(T)$^a$ & 1.731 & 2.49 & 792 \\
\hline
exp.\footnote{Ref.\ [16]} & 1.749 & 2.61$\pm$0.22 & 785 \\
\hline
\end{tabular}
\end{minipage}
\end{center}
\end{table}

\newpage
\begin{table}
\begin{center}
\begin{minipage}{5in}
\caption{\label{4}Intra-ionic correlation-energy increments and total
contribution,
$\Delta$E$_{\rm coh}$(MgO), of these increments to the cohesive energy of bulk
MgO (in atomic units, 1 a.u.\ =
27.2114 eV).}
\vspace{5mm}
\begin{tabular}{|c|rrr|c|}
\hline
&ACPF &CCSD& CCSD(T)&
\multicolumn{1}{|c|}{Weight\footnote{Weight factor in the incremental expansion
of the bulk correlation energy
(per primitive unit cell) of MgO.}} \\
\hline
  \multicolumn{1}{|c|}{Mg$\rightarrow$Mg$^{2+}$ \footnote{($4s4p$)-valence
basis set, cf.\ Table \ref{2}. PP+CPP
  description of Mg$^{2+}$.}}
  & 0.04690 & 0.04690 & 0.04690 & 1 \\
  \multicolumn{1}{|c|}{O$\rightarrow$ O$^{2-}$
 \footnote{[$5s4p3d2f$] basis set, cf.\ Table \ref{1}. Stabilization of
O$^{2-}$ by
 means of 6 Mg$^{2+}$ pseudopotentials at the nearest-neighbour positions of
the MgO crystal ($r_{\rm MgO}$=2.105 \AA).
 Madelung field of outer ions in 7x7x7 cube represented by point charges
$\pm$2.}}
 & -0.09548 & -0.09646 & -0.10162 & 1 \\
\hline
 \multicolumn{1}{|c|}{$\Delta$E$_{\rm coh}$(MgO)\footnote{Correlation-energy
change per primitive unit cell,
 for separation into neutral
 ground-state atoms.}}
 & 0.04858  & 0.04956 & 0.05472 & - \\
\hline
\end{tabular}
\end{minipage}
\end{center}
\end{table}

\newpage
\begin{table}
\begin{center}
\begin{minipage}{5in}
\caption{\label{5}Inter-ionic two-body correlation-energy increments and total
contribution, $\Delta$E$_{\rm
coh}$(MgO), of these increments to the cohesive energy of bulk MgO (in a.u.).
The notation $A$-$B \rightarrow n$
means that the increment describes a $n$th nearest neighbour pair of $A$ and
$B$ ions in the
crystal.}
\vspace{5mm}
\begin{tabular}{|c|rrr|c|}
\hline
 & \multicolumn{1}{|c}{ACPF\footnote{For the
definition of the increments, see Eq.\ 2.}$^,$\footnote{The ($4s4p$)
valence basis set was used for Mg (Table \ref{2}) and the [$5s4p3d2f$] basis
for O (Table \ref{1}).}
 } & \multicolumn{1}{c}{CCSD$^{a,b}$} & \multicolumn{1}{c}{CCSD(T)$^{a,b}$}
 & \multicolumn{1}{|c|}{weight\footnote{Weight factor in the incremental
expansion of the bulk correlation
  energy (per primitive unit cell) of MgO.}} \\
\hline
Mg-O $\rightarrow 1$ \footnote{O$^{2-}$ at (0,0,0) with 6 surrounding Mg$^{2+}$
carrying a PP (one of them an
additional CPP) in a 7x7x7 cube of point charges.}
& -0.003031 & -0.003040 & -0.003037 & 6 \\
Mg-O $\rightarrow 2$ \footnote{Mg$^{2+}$ carrying a PP+CPP at (0,0,0), O$^{2-}$
at (1,1,1) with a cage of
6 surrounding Mg$^{2+}$ PP, in a 7x7x7 cube of point charges.}
& -0.000082 & -0.000083 & -0.000084 & 8 \\
Mg-O $\rightarrow 3$ \footnote{same as e), but  Mg$^{2+}$ at (-1,0,0) and
O$^{2-}$ at (1,1,0).}
& -0.000014 & -0.000013 & -0.000014 & 24 \\
\hline
O-O $\rightarrow 1$ \footnote{O$^{2-}$ ions at (0,0,0) and (0,1,1), 10
neighbouring Mg$^{2+}$ carrying a PP,
remaining ions of a
7x8x8 cube represented by point charges.}& -0.002740 & -0.002538 & -0.002973 &
6 \\
O-O $\rightarrow 2$ \footnote{O$^{2-}$ ions at ($\pm1$,0,0), 11 Mg$^{2+}$ PP,
in a cube of 9x9x9 ions.}
& -0.000238 & -0.000225 & -0.000264 & 3 \\
O-O $\rightarrow 3$ \footnote{O$^{2-}$ ions at (0,0,0) and (2,1,1), 12
Mg$^{2+}$ PP,
in a cube of 9x9x9 ions.} & -0.000066 & -0.000061 & -0.000072 & 12 \\
O-O $\rightarrow 4$ \footnote{O$^{2-}$ ions at (-1,-1,0) and (1,1,0), 12
Mg$^{2+}$ PP,
in a cube of 9x9x9 ions.} & -0.000027 & -0.000024 & -0.000029 & 6 \\
O-O $\rightarrow 5$ \footnote{O$^{2-}$ ions at (-2,0,0) and (1,1,0), 12
Mg$^{2+}$ PP,
in a cube of 9x9x9 ions.} & -0.000014 & -0.000012 & -0.000014 & 12 \\
\hline
$\Delta$E$_{\rm coh}$(MgO) & 0.037454 & 0.036139 & 0.039066& -\\
\hline
\end{tabular}
\end{minipage}
\end{center}
\end{table}

\newpage
\begin{table}
\begin{center}
\begin{minipage}{5in}
\caption{\label{6}Inter-ionic three-body correlation-energy increments and
total contribution, $\Delta$E$_{\rm
coh}$(MgO), of these increments to the cohesive energy of bulk MgO (in a.u.).
The notation
$A$-$B$-$C \rightarrow n$ means that $AB$ and $BC$ are nearest neighbours of
their species, while $AC$ are $n^{th}$
neighbours.}
\vspace{5mm}
\begin{tabular}{|c|rrr|c|}
\hline
 & \multicolumn{1}{|c}{ACPF\footnote{For the
definition of the increments, see Eq.\ 3.}$^,$\footnote{The ($4s4p$)
valence basis set was  used for Mg (Table \ref{2}) and the [$5s4p3d2f$] basis
for O (Table \ref{1}).}
 } & \multicolumn{1}{c}{CCSD$^{a,b}$} & \multicolumn{1}{c}{CCSD(T)$^{a,b}$} &
 \multicolumn{1}{|c|}{weight\footnote{Weight factor in the incremental
expansion of the bulk correlation
  energy (per primitive unit cell) of MgO.}} \\
\hline
O-O-O $\rightarrow 1$ \footnote{O$^{2-}$ ions at positions (1,0,0), (0,1,0),
(0,0,1), surrounded by a cage of
Mg$^{2+}$ PP each, within a cube of 7x7x7 ions, with point charges $\pm$2
simulating the outer ions.} &
 0.000069 & 0.000063 & 0.000067 & 8 \\
O-O-O $\rightarrow 2$ \footnote{Same as d), but O$^{2-}$ positions (1,0,0),
(-1,0,0), (0,0,1).} &
 0.000014 & 0.000016 & 0.000013 & 12 \\
\hline
O-Mg-O $\rightarrow 1$ \footnote{Mg$^{2+}$ PP+CPP at (0,1,0), O$^{2-}$ ions
surrounded by Mg$^{2+}$ cage at
(0,0,0) and (0,1,1) in a  7x8x8 cube of point charges.}
& 0.000013&0.000014&0.000016&12\\
O-Mg-O $\rightarrow 2$ \footnote{Mg$^{2+}$ PP+CPP at (0,0,0), O$^{2-}$ ions
surrounded by Mg$^{2+}$ cage at
(1,0,0) and (-1,0,0) in a  7x7x7 cube of point charges.}
& -0.000023&-0.000022&-0.000026&3\\
\hline
$\Delta$E$_{\rm coh}$(MgO) & -0.000807&-0.000798&-0.000806 &- \\
\hline
\end{tabular}
\end{minipage}
\end{center}
\end{table}

\newpage
\begin{table}
\begin{center}
\begin{minipage}{5in}
\caption{\label{7}Total correlation contribution, $\Delta$E$_{\rm coh}$(MgO),
to the bulk cohesive energy per primitive unit cell
of MgO with respect to separated neutral atoms, from an expansion using the
local
increments of Tables 4 through 6, compared to the 'experimental' value. All
data in a.u.. }
\vspace{5mm}
\begin{tabular}{|c|lr|}
\hline
& \multicolumn{2}{|c|}{$\Delta$E$_{\rm coh}$(MgO)} \\
\hline
ACPF& 0.0852 & (79\%) \\
CCSD& 0.0849 & (79\%) \\
CCSD(T) & 0.0930 & (86\%) \\
\hline
exptl.\footnote{Difference of the experimental [19] and the HF value [6c] of
the
MgO cohesive energy; the experimental value has
been corrected for zero-point energy.}
& 0.1079 &  \\
c-DFT\footnote{Gradient-corrected correlation-energy density functional, Ref.\
[6b].} & 0.087 &  \\
\hline
\end{tabular}
\end{minipage}
\end{center}
\end{table}

\end{document}